\documentstyle{article}

\def\ben{\begin{equation}}
\def\een{\end{equation}}
\def\bea{\begin{eqnarray}}
\def\eea{\end{eqnarray}}
\input amssym.def
\input amssym.tex
\begin{document}

\hfuzz=100pt
\title{Wrapping Branes in Space and Time}
\author{G. W. Gibbons
\\
D.A.M.T.P.,
\\ Cambridge University, 
\\ Silver Street,
\\ Cambridge CB3 9EW,
 \\ U.K.}
\maketitle

\begin{abstract}
Branes may be approximated semi-classically by 
solutions of supergravity theories with event and Cauchy horizons.
I suggest  that if one wishes to avoid singularities
and to capture accurately some of the properties
of branes then these classical spacetimes must be identified
so as to render them periodic in time. 

\end{abstract}

\section{Introduction}

Great progress in string and M-theory  has been achieved in recent years by  
considering classical solutions of their low energy supergravity limits.  
The solutions of greatest interest typically
represent electrically  or magnetically charged extreme p-branes
with degenerate (i.e zero-temperature)  event horizons.  These $(p+q+4)$
dimensional metrics may be expressed in isotropic form as 
$$
ds^2 = H^{ -{ 1 \over p+1}} \{ -dt^2 + d{\bf x }_p. d{\bf x}_p \} + 
H^{ 1\over q+1}  d{\bf y}_{q +3}.d{\bf y}_{q +3}
$$
where $H$ is a  harmonic functions on the transverse euclidean space ${\Bbb E}^{q +3}$ with coordinates ${\bf y}_{q +3}$.  In addition one has a $p+2$ form or 
$q +1$ form field strength and one may have a dilaton and other moduli fields.
One of the simplifications of M-theory is that the dilaton 
and other scalars are absent. In what follows I shall mainly restrict
myself to the M-theory case for which $p=2$ ( the membrane) and $p=5$,
the five-brane. 
 
As explained in a recent paper \cite{G2} the coordinates
$(t, {\bf x}_p, {\bf y}_{q+3})$ provide a harmonic map of the
portion of full spacetime  exterior to the brane into
${\Bbb R}^{p+q+4}$. The horizons then get mapped into distributional
sources. That description is very close
to one based on elementary or fundamental branes 
moving in flat or almost flat spacetime
but it
leaves out all the subtle geometrical details
connected with the  the horizons and their interiors. In this paper
by contrast It is precisely  some aspects of the horizon geometry
taht I wish to explore. To some extent one is dealing with 
some complementary
features of branes.

\noindent The electric or magnetic charges carried by the $p+2$ form or $q+2$ form respectively are central charges and saturate
the supergravity version of the familiar Bogomol'nyi bound 
of magnetic monopole theory. 
This implies that the solutions are supersymmetric and  one anticipates,
at least in those cases when only one half of the maximum supersymmetries ar 
unbroken, 
that some  classical properties may be extrapolated to the fully 
quantum regime.
      
\noindent These self-gravitating solutions are similar in many ways to
their analogues
in theories without gravity. For example they may be considered to interpolate 
spatially between different maximally supersymmetric vacua \cite{GT} \cite{DGT}.
For the M-2-brane the near horizon state corresponds
to  the $AdS_4 \times S^7$. For the M-5-brane it corresponds to 
its dual: $AdS_7\times S^4$. 

\noindent The electrical solutions resemble the Coulomb solutions 
(for which ${\bf E}^a = {\cal D} {\bf \Phi}^a$),  
of non-Abelian Yang-Mills theory have singularities inside the event horion.
One expects them to provide a good approximation
to the field generated by many \lq fundamental \rq or \lq elementary\rq  $p$-branes sitting on top of one another.
By contrast  the magnetically charged solutions resemble the solitons 
(for which ${\bf B}^a = {\cal D} {\bf \Phi}^a$) of non-Abelian Yang-Mills theory.
In some cases, such as the 5-brane of M-theory, they are completely non-singular.

\noindent On the horizon $H^{-1}$ vanishes and thus if $p$ or is greater than zero the area of the event horizons  vanish. In other words fundamental
and solitonic extreme branes carry no Bekenstein-Hawking entropy.
This is what one usually expects expects of fundamental objects and solitons
whose fundamental dynamics  is non-dissipative.
Of course 0-branes, even extreme ones,
do have entropy but this is associated with the
"intersection" of branes.

\noindent However there remain some difficulties with this interpretation.
They include:

\begin{itemize} 
\item The solutions above represent infinite p-branes.
One wants to be able to wrap them over tori, 
but this cannot be done naively because, as we shall see shortly,
translations of the world volume coordinates ${\bf x}_p$ do not
act freely on the horizon.

\item M-theory seems to admit 2-branes whose world volumes
are not space-orientable. Five-brane world volumes however 
must be orientable. How do we construct them?

\item The existence of horizons,
albeit extreme, would seem on the face it to lead
semi-classically to "information loss" and dissipative dynamics.

\item The existence of Cauchy horizons 
allows passge through to an infinite chain of other universes.
This is surely rather a superfluous if one wants to describe a single brane.
\end{itemize}

\noindent Of course one might argue that these are just symptoms of 
a breakdown of the the semi-classical approximation. However since
M-theory has no dimensionless parameter to 
expand in this is a rather dangerous argument since a failure to capture
the essential qualitative features of M-theory,
such as toroidal and possibly non-orientable world volumes
in the classical limit
would cast doubt on the relevance of the classical solutions at all. 
Moreover recent work in M-theory has revealed that
a surprisingly large amount of the structure is reflected
in the classical solutions see e.g. \cite{G1}.
Therefore in this paper I shall take a different approach.
I will explore a possible resolution
 by considering how one can identify 
the classical solutions by isometries without fixed points
and thus  avoid the introduction of extra singularities.
This leads to the suggestion that the solutions should be identified
periodically in time. It has previously been argued
that an approximation to the world volume theories
may be obtained by considering the singleton and doubleton representations
of the relevant Anti-De-Sitter groups \cite{GT}. I shall argue that
my suggestion is consistent with that viewpoint.

\section{ Mapping into  $CAdS_{p+2} \times S^{q +2}$}.

A convenient way of exhibiting the global 
structure of the single extreme $p$-brane
spacetimes is to map them onto a portion of the universal covering space
of Anti-de-Sitter spacetime times the $(q +2)$-sphere \cite{GHT},
${C {AdS_{p+2}}} \times S^{q +2}$. For multi-$p$-brane
solutions many such patches will be required. We begin by 
reviewing some standard material on $AdS_{p+2}$.

\subsection{ Some properties of $AdS _{p+2}$ and $CAdS_{p+2}$}

One usually defines $AdS_{p+2} $ as the hyperboloid
 in ${\Bbb E}^{p+1,2}$ whose coordinates are  $X^A$, $A=0, \dots {p+3}$
and which is given by
$$
(X^0)^2 - (X^i)^2   + (X^{p+2})^2 =1,
$$
with its induced lorentzian metric, where $i=1,\dots,p+1$. Note that $X^0$ and $X^{p+2}$ are timelike coordinates and $X^i$ are spacelike coordinates.
$AdS_{p+2}$ is not simply connected, 
it has topology $S^1 \times {\Bbb R}^{p+1}$. The simply connected
universal covering space has topology ${\Bbb R}^{p+2}$ and is traditionally called ${C {AdS_{p+2}}}$. 
We introduce coordinates $(\tau,\chi, {\bf n}) $ making manifest the maximal compact subgroup
$SO(p+1) \times SO(2) \subset SO(p+2,2)$
 by
$$ 
X^0 =\cos \tau \cosh \chi,
$$
$$
X^{p+2}= \sin \tau \cosh \chi,   
$$
$$
X^i= \sinh \chi {n^i}.
$$
where $n^i$ is a unit vector in ${\Bbb E}^{p+1}$ defining  $S^p$.
In these coordinates the metric takes the globally static form:
$$
ds^2 = -\cosh ^2 \chi d \tau^2 + d \chi ^2 + \sinh \chi d \Omega ^2 _{p}
$$
where $d\Omega ^2 _p$  is the standard round metric on $S^p$.
We take $0\le \tau \le 2 \pi$
to get $AdS_{p+2}$ but allow it to take all real values to obtain $CAdS_{p+2}$. Both $AdS_{p+2}$ and $CAdS_{p+2}$ are space and time orientable. Evidently $AdS_{p+2}$ is periodic in time and has closed timelike curves. In fact every timelike geodesic is periodic with proper-time
period $2\pi$.
The universal covering space $CAdS_{p+2}$ is not periodic in time and has no closed timelike curves. 

The antipodal map $J$ lies in the centre of $O(p+2,2)$
and acts on $AdS_{p+2}$ by 
$$
J: X^A \rightarrow -X^A.
$$
In global, static coordinates this becomes $(\tau, \chi, {\bf n}) \rightarrow
(\tau +\pi, \chi, -{\bf n})$. The antipodal map preserves time orientation
but, since it induces the antipodal map on $S^p$, it reverses space orientation
or preserves it depending upon whether $p$ is even or odd respectively. 
In fact $J$ lies in the identity component of $SO(p+1,2)$
if $p$ is odd. On $AdS_{p+1}$
$J$ acts  an involution : $J^2={\rm id}$. We may extend $J$ to an
action of the integers  on $CAdS_{p+2}$ by defining
$$
J^n:(\tau, \chi, {\bf n}) \rightarrow
(\tau +n \pi, \chi, (-1)^n{\bf n}). 
$$
 Clearly we can take the quotient of $CAdS_{p+2}$
by any integer power of $J$. If we take $J^{2k}$, $ k \in {\Bbb Z}$
 we get the $k$-fold cover of
$AdS_{p+2}$.  Of course it is now a $k$-fold cover of 
$SO(p+1,2)$ which acts.

\subsection {Horospheric coordinates}

For brane purposes it is convenient to 
introduce another set of coordinates for $AdS_{p+2}$.
They are $(t, {\bf x}_p, z)$ where ${\bf x}_p$ has $p$ components
and make manifest the action of the
Poincare subgroup of $SO(p+1,2)$.
One sets
$$
X^0= {t \over z},
$$
$$
X^a = {x^a \over z},
$$
$$
X^{p+2}  - X^{p+1} = { 1\over z},
$$
and
$$
X^{p+2} + X^{p+1} = z + { ( {\bf x} _p.{\bf x}_p -t^2) \over z}.
$$
The metric takes the form
$$
ds ^2 = { 1 \over z^2} \Bigl \{ -dt^2 + d{\bf x}_p.d{\bf x}_p + dz^2 \Bigr \}.
$$
These coordinates provide a foliation of $AdS_{p+2}$ by
flat timelike hypersurfaces $z={\rm constant}$. These hypersurfaces
are the intersections of the null hyperplane with the hyperboloid.
In fact exact horosphere is a solution of the Dirac action for
a test p-brane and so we see very vividly that
the foliation by
horospheres corresponds to the idea of a supergravity 
p-brane consisting  of a very large number of
fundamental p-branes. 

Horospheric coordinates cover only one half of $AdS_{p+1}$.
They break down at $z=\infty$ which is the intersection of the null 
hyperplane which passes through the origin,
$$
X^{p+2}  - X^{p+1} = 0,
$$
with the hyperboloid. The antipodal map $J$ simply reverses the sign of $z$. If one 
were to identify $AdS_{p+2}$ under the action of $J$ 
a single horospheric patch would suffice. 
If one passes to the universal covering space $CAdS _{p+2}$
one needs infinitely many patches. One may regard 
$z=0$ as a degenerate Killing horizon associated to the  
every where non-spacelike  time translation Killing vector 
field $\partial \over \partial t$. The Killing vector $\partial \over \partial t$ generates the action $t \rightarrow t + c$ 
and these act freely on $AdS_{p+2}$. To see this one may use the embedding coordinates. time translations act as   
$$
X^0 \rightarrow x^0 + c (X^{p+2} - X^{p+1}),
$$
$$
X^i \rightarrow X^i,
$$
$$
X^{p+2} - X^{p+1} \rightarrow X^{p+2} - X^{p+1},
$$
and
$$
X^{p+2} + X^{p+1} \rightarrow X^{p+2} + X^{p+1}+ 2c X^0 + c^2 (X^{p+2} - X^{p+1}).
$$
Any possible fixed points must satisfy
$$
X^{p+2} - X^{p+1}=0,
$$
and 
$$
X^0=0.
$$
But this implies that
$$
-X^i X^i =1
$$ which is impossible.

By contrast the spatial translations ${\bf x}_p \rightarrow {\bf x}_p + {\bf a}_p$ do not act freely. The action on the embedding coordinates is
$$
X^0 \rightarrow X^0,
$$
$$
X^i \rightarrow X^i + a^i (X^{p+2} - X^{p+1}),
$$
$$
X^{p+2} - X^{p+1} \rightarrow X^{p+2} - X^{p+1}
$$
and 
$$
X^{p+2} + X^{p+1}\rightarrow X^{p+2} +X^{p+1} -2 a^i X^i - a^2 (X^{p+2} - X^{p+1}).
$$
The fixed points lie on the horizon
$$
X^{p+2} - X^{p+1}=0
$$
and must also satisfy
$$
a^i X^i=0.
$$   
It is easy to satisfy this condition.

It follows form this that one may, if one wishes,
identify  $AdS_{p+2}$ under a discrete time translation and obtain a 
non-singular quotient but one cannot identify under a discrete 
space translation and obtain a non-singular quotient.
It is also clear that spatial parity $\Omega_p$ ,
i.e, ${\bf x}_p \rightarrow - {\bf x}$ does not act freely. 
In the brane context  there is no good motivation for identifying under
world sheet time translations but if the world sheet is toroidal one must
 identify under a discrete group $\Lambda_p$ of spatial translations
and if one considers orientifolds, or if one wishes to consider 
non-orientable world sheets one is interested in identifying under spatial 
parity $\Omega_p$. Note that as a general principle one should not 
identify under
time orientation reversing isometries, either in spacetime or 
on the world volume since that leads to  difficulties with quantization.
One would be  forced into real quantum mechanics.

\subsection{ Maximal extensions of $p$-branes.}

Having developed the necessary material about $AdS_{p+2}$ 
we are in a position to discuss the maximal extensions of the
$p$-brane solutions. The basic idea is to map them 
into the universal cover, $CAdS_{p+2}$,
pushing forward the metric. Note that the metric will {\sl not} be conformal
to the $AdS_{p+2}$ metric.

We may take,, by choosing the unit of length appropriately, 
$$ H= \bigl (1+ {1 \over \rho^{q+1}} \bigl )= 
{ 1 \over \bigl (1-{1 \over r^{q +1} } \bigr )},
$$
where $\rho =\sqrt {y^2}$ is an isotropic radial coordinate and $r$ 
is a Schwarzschild radial coordinate. The $(q+1)$- volume of
a transverse $q+1$-sphere is thus $r^{q + 1}  \omega _{q+1} $ where  $\omega _{q+1}$ is the volume of a unit
$(q +1)$-sphere.  
The required mapping is obtained by setting
$$
{p+1\over z} =H^{- {1 \over p+1}}
$$
The horizon, at which $H$ diverges, is mapped to the degenerate horizon at $z=\infty$ and in the neighbourhood of the horizon we have the {\sl asymptotic form}
of the metric
$$ 
ds^2\approx  \big ({p+1\over z} \bigr )^2  \Bigl \{ -dt^2 + d{\bf x}_p.d{\bf x}_p + dz^2 \Bigr \} + d\Omega^2 _{ q+2}.
$$
The exact form of the metric is  obtained by expressing $\rho$ in terms of $z$ and substituting. Note that the $AdS_{p+2}$ factor is scaled up by a factor $p+1$. The maximal extension is obtained by lifting this metric
to the universal cover $CAdS_{p+2}$. 
The result depends upon the parity of $p$. If $p$ is even 
the spacetime is like extreme Reissner-Nordstrom :there
is an infinite chain of asymptotically flat regions joined via degenerate
horizons to an infinite chain of internal regions containing timelike
singularities at $r=0$ which are inside the horizons. By contrast if
$p$ is odd the solution is symmetrical and completely non-singular:
the extension consists of two infinite chains
of asymptotically flat regions joined by degenerate horizons.

We may express this in terms of the map $J$ which continues to act 
freely on the spacetime manifold but not necessarily as an  isometry. 
Thus if
$p$ is even
$J$ takes exterior regions to interior regions but  is {\sl not} an 
isometry
of the $p$-brane metric. However if $p$ is odd $J$  {\sl is}
an isometry of the $p$-brane metric and takes one exterior region to an 
adjacent one. Of course $J^2$ always acts by isometries and takes 
one along the chains of exterior or interior regions.

It follows from this that one may identify the $p$-brane metric by any even
integer power of $J$ if $p$ is odd even and any integer power of $J^2$
if $p$ is even without introducing any more singularities than were 
originally present. The resulting spacetimes will be time orientable
but contain closed timelike curves. Taking the smallest possible case this means
that for the 2-brane one has just one external region and one 
internal region.  A timelike curve  falling through the future horizon
from the outside may either hit the singularity or leave the internal region and return to the external region through the past horizon before  it left it.
In the case of the 5-brane the curve would never leave 
the external region since just as it crossed the future horizon it would 
re-appear entering the exterior region through the past horizon. More elaborate delay phenomena may
can take place if one identifies under higher powers of $J$.
In many ways the nicest case is the minimal one because in that case all
but one infinite region is eliminated. 
This means that there is no possibility of information being lost from one
infinite region into another. Any possible information loss could only be
into the singularity.
 However  that is where the semi-classical approximation breaks down 
and so one might argue that one should not be concerned about it since it might be just an artefact of an inapplicable approximation. By contrast the
loss of information to other infinite region is is potentially more worrying
because the loss is  via regions near the horizon
in which the curvature need not be especially high.

\subsection {Toroidal p-branes.}

It is clear form our discussion of $AdS_p$ that one cannot identify 
under the action
of
the discrete group of $p$ spatial translations $\Lambda$ to obtain a 
non-singular quotient since the lattice $\Lambda_p$ (in the mathematical 
sense
of a finite freely generated abelian group)
does not act freely. For the same reason we cannot identify the $p$-brane
spacetimes under the action of the lattice $\Lambda_p$ to obtain
a regular solution with a toroidal  world volume. In other words
we cannot obtain a regular non-singular classical solution representing 
a $p$-brane wrapped around a toroidal cycle. Since much of current
work on D-branes assumes that this is possible this is rather disturbing.
There is, however, an obvious  way out. One may compose the lattice $\Lambda$
with any power of the antipodal map $J$.
in other words the  map $J^n \cdot\Lambda$ does act freely on $CAdS_{p+2}$. Thus as long as one wraps the $p$- brane in time direction as well as in space one will obtain a smooth quotient. The integer $n$ should be multiple
of 2 for the 2-brane but it could be any integer for the 5-brane.
We shall see in the next section that singleton considerations suggest
that we should take $n=4$ for the 2-brane and $n=1$ for the 5-brane.   

\subsection {Orientifolding and non-orientable world 
volumes.}

The previous remarks can be immediately extended
to cover  world volumes which are not space-orientable.
World volume parity $\Omega_p$   we take to be
$$
\Omega_p:  {\bf x}_p \rightarrow  -{\bf x}_p.
$$
in terms of the embedding coordinates this takes
$$
X^i \rightarrow -X^i
$$
and leaves the other coordinates unchanged. Clearly $\Omega_p$
does not act freely but we could compose with  with any
integer  power
of $J$
to get something with no fixed points.

\noindent Suppose we wanted a Klein Bottle world volume. 
The Klein bottle may obtained by identifying ${\Bbb E}^2 $ in say the 1-direction 
direction by a translation and in the 2-direction by a glide reflection,
in other words other by the composition of a 
translation in the 2-direction and a reflection in the 2-axis.
As before this does not act freely at the horizon but composing 
with an appropriate integer power of $J$ we acn obtain a free action.
Thus agin we are led to wrap in time as well as in space.

\subsection {Intersecting branes and black holes}

There exist solutions of M-theory depending upon more than one
harmonic function which represent \lq intersecting \rq branes.
Not all such solutions are non-singular but some are.
Thus there is a solution  depending on three harmonic functions
$H_1$, $H_2$, $H_3$ on ${\Bbb E}^4$ with coordinates ${\bf y}_4$  
 which may be interpreted as three 2-branes
\lq intersecting \rq on a point. 

The metric takes the form
$$
ds^2 = -(H_1 H_2 H_3)^ { -{ 2\over 3} } dt^2  + (H_1 H_2H_3 )^{ 1\over 3}  d{\bf y} _4.d{\bf y}_4
$$ 
$$
+ ( {H_2 H_3 \over H _1^2} ) ^{ 1 \over 3} (dx_1 ^2  + dx_4^2 ) 
+( {H_3 H_1 \over H _2^2} ) ^{ 1 \over 3}( dx_2 ^2  + dx_5^2 ) 
+( {H_1 H_2 \over H _3^2} ) ^{ 1 \over 3} (dx_3^2 + dx_6^2) .
$$
The $6$ Killing fields $\partial \over \partial x_i $ , $i=1,2,3,4,5,6$
never vanish and so one  may identify the coordinates $x_i$, $i=1,2,3,4,5,6$
without introducing any quotient singularities. With this identification
 the metrics admit a free 
action of the torus group $T^6$. They may therefore be  
dimensionally reduced to $4+1$ spacetime dimensions. The resulting  solution
represents a black hole. 
We may further reduce to $4+1$ dimensional  solution 
representing an extreme black hole.

Near the throat the geometry behaves like $AdS_2 \times S^3$.
The causal structure is Reissner-Nordstrom-like. The maximal
extension is obtained using the same procedure as described above.  
It follows that one may extend the action of the antipodal map for
two-dimensional Anti-de-Sitter space  $J$ to the maximal extension
such that $J$ interchanges the interior with the exterior and 
so that $J^2$
acts isometrically. Thus we may again identify periodically, 
just as we may for
the spacetime of a single 2-brane.

Another non-singular solution representing the 
intersection of three 5-branes over 
a string and depending upon three harmonic function on ${\Bbb E}^3$
 has the metric
$$
ds^2 = (H_1 H_2H_3)^{-{1 \over 3}} ( -dt^2 + dx_7^2) + 
(H_1 H_2 H_3)^{ 2 \over 3} d{\bf y}_3 .d {\bf y}_3 $$ $$
+ ({H_1^2 \over H_2 H_3}) ^{ 1 \over 3} (dx_1^2 + dx_4^3 )
+ ({H_2 ^2 \over H_3 H_1}) ^{ 1 \over 3} (dx_2^2 + dx_5^3 )
+({H_3 ^2 \over H_1 H_2}) ^{ 1 \over 3} (dx_3^2 + dx_6^3 )
$$

Again there is a free $T^6$ 
action and one may dimensionally reduce to get a solution in 
$4+1$ dimensions. The solution is independent of 
the coordinate $x_7$ and thus represents an extreme string. 
Near the throat it looks like $AdS_3 \times S^2$.
However the Killing field $\partial \over \partial x_7$
does not generate a free action and further reduction 
dimensions, to get 3+1 dimensional black hole solutions is 
problematical.
Let us therefore  remain in 4+1 dimensions and maximally extend. 
In the simples case in which we assume that 
$H_1=H_2=H_3 = 1 + { 1\over \rho}$
may apply the results of [2] to discover we find that 
maximal 
extension is symmetric and singularity free like the 5-branes 
whose intersection it is.
The extension with three distinct
harmonic functions $H_i$ each of the form:
$$
H_i= 1 + { \mu_i \over \rho}, 
$$
is only slightly more difficult.
One sets
$$
H_1H_2H_3 = \mu_i\mu_2\mu_3 ({ z^2 \over 2} )^3.
$$
It is clear that the metric functions will be even functions of $z$
and hence the antipodal map $J$ will act isometrically.  

Now if we wish to the string coordinate identify $x_7$ we may do so
as long as we compose with $J$ as well.  

\section{ Singletons and Doubletons}

Some extra evidence in favour of the identifications I am proposing
is 
provided by considering the singleton and doubleton representations
of $SO(3,2)$ and $SO(6,2)$ and their super-extensions.
These have no analogues among the representations of the Poincare groups
of $4$ or $7$ dimensional Minkowski spacetime. They do not, therefore, 
survive in the limit as the curvature of $AdS_4$ or $AdS_7$ is sent
 to zero. Even if the curvature is non-vanishing they
cannot be realized as quantum fields propagating in the bulk
in $AdS_4$ or $AdS_7$, respectively.  However they can be realized as
conformally-invariant supersymmetric quantum field theories
on $S^1 \times S^2$ or $S^1 \times S^5$. 
By defining a coordinate $\lambda$ by  
$$
\sin \lambda= \tanh \chi
$$
one may conformally embed $CAdS_{p+2}$ into the static
Einstein universe. The metric becomes
$$
ds^2 = { 1 \over \cos^2 \lambda} \Bigl \{ -d \tau^2 + d \lambda ^2 + \sin^2 \rho 
d \Omega ^2 _p \Bigr \}.
$$
The conformal boundary of $AdS_{p+2}$ is given by $\lambda = {\pi \over 2}$.
Thus one may regard $S^1 \times S^p$ as the conformal boundary of
$AdS_{p+2}$ and the isometry group $SO(p+1,2)$ 
of $AdS_{p+2}$ acts by conformal transformations
on  $S^1\times S^p$. If $p=2$ or $p=5$
These quantum field theories
 have the same degrees of freedom as the world volume fields of $p$-branes 
and it was suggested that  the lowest scalar component, call it $X$,
may be regarded as the transverse oscillation of the $p$-brane which is
localized in the vicinity of the horizon. Note that this interpretation is 
similar to, but not exactly the same as, the idea of the \lq membrane at the
end of the universe \lq since we are now  regarding a small timelike
tube around the horizon as the location of the 2-brane or 5-brane
and claiming that the membrane oscillations can be desribed group-theoretically in terms of singletons. Presumably there are fluctuations 
of the spacetime geometry carrying the singleton representations.

In the linearized 
approximation we are considering here $X$ satisfies the conformally
invariant wave equation on $S^1 \times S^p$
$$
-\nabla ^2 X  + {p-1 \over 4p}  R X=0.
$$
where $R$ is the Ricci scalar of $S^1 \times S^p$ which equals the Ricci scalar of $S^p$. On a unit $p$-sphere $R= p(p-1)$, and the eigenvalues of the
spherical harmonics$ Y_l({\bf n}) $ on $S^p$ are $l(l+p-1)$, where $l=0, 1,\dots $. Thus  the modes of $X$ behave like
$$
X_l= \exp{-i(l+{ p- 1 \over 2})\tau }  Y_l({\bf n}).
$$   

Thus 
$$
X(t+2\pi) = (-1)^ {p-1} X(t).
$$
 
It follows that if $p$ is odd, as it is for the 5-brane, then $X$ is periodic 
with the basic Anti-de-Sitter  period. If however $p$ is even, as it is for the
2-brane, then $X$ is periodic with half the basic  Anti-de-Sitter period.
In other words the theory is really defined not on $S^1 \times S^p$ 
but its double cover.

We may express this in terms of the antipodal map $J$ which acts on
the boundary of $AdS_{p+2}$ as 
$$
J: (\tau,{\bf n}) \rightarrow (t+\pi, -{\bf n}).
$$
In other words we advance $t$ by half a period and compose with 
the antipodal map on $S^p$. As before we may extend
$J$ to act in the obvious way on the universal covering space
${\Bbb R} \times S^p$. Now the spherical harmonics satisfy
$$
Y_l(-{\bf n}) = (-1)^l Y_l({\bf n}).  
$$
Thus 
$$
X \cdot J =  i^{p-1} X.
$$
  
The singletons are therefore invariant under $J$
if $p=5$  and under $J^4$ if $p=2$.
thus indeed the singletons indicate that one may 
identify the inside with the outside of the 5-brane.
In the case of the 2-brane however they indicate that one needs at least two
exterior and two interior regions.  

The above analysis of singletons would also apply in the case of 
strings, i.e. $p=1$. It is striking that we get the same result
as for the 5-brane. Particularly so when we recall that we may construct
a string as a triple intersection of 5-branes.

\section{Conclusion}

In this paper I  have studied the global structure of the classical
 spacetimes of 
 branes and their non-singular intersection in  M-theory. 
I have  found that if one  wraps the branes in space it is 
also necessary to wrap them in time in order avoid quotient singularities.
This procedure is consistent with properties of the relevant
singleton
representations and may  help resolve some of the puzzles connected with
the possible loss of information in spacetimes with horizons and eliminates
the infinite chain of other universe. 

An initial draft of  the present paper was written 
over a year ago,
long before the current rise in the number of papers
relating p-branes and
singletons. It has been revised
recently but since the ideas expressed here, while being relevant to that
activity are independent of it, I have made no attempt
to include them in the bibliography.

\section{Acknowledgements}

It is a pleasant duty to acknowledge helpful conversations
with Andrew Chamblin, Stephen Hawking, George Papadopoulos,
Kelly Stelle  and Paul Townsend.

\end{document}